\documentclass[preprint,aps,epsfig,floats,showpacs]{revtex4}
\usepackage{graphicx}

\begin{document}
%\draft

\title{Superconducting properties of Nb thin films deposited on porous silicon templates}
\author{M. Trezza$^1$, S. L. Prischepa$^2$, C. Cirillo$^1$, R. Fittipaldi$^1$, M. Sarno$^{3,4}$,
D. Sannino$^{3,4}$, P. Ciambelli$^{3,4}$, M. B. S.
Hesselberth$^5$, S. K. Lazarouk$^2$, A. V. Dolbik$^2$, V. E.
Borisenko$^2$, and C. Attanasio$^{1,4}$}

\address{
$^1$ Laboratorio Regionale SuperMat, CNR-INFM Salerno and
Dipartimento di Fisica \lq\lq E. R. Caianiello\rq\rq,
Universit\`{a} degli Studi di Salerno, Baronissi (Sa) I-84081,
Italy\\
$^2$ State University of Informatics and RadioElectronics, P. Brovka street 6, Minsk 220013, Belarus\\
$^3$ Dipartimento di Ingegneria Chimica e Alimentare,
Universit\`{a} degli Studi
di Salerno, Fisciano (Sa) I-84084, Italy\\
$^4$ NANO\_ MATES,  Research Centre for NANOMAterials and
nanoTEchnology at Salerno University, Universit\`{a} degli Studi
di Salerno, Fisciano (Sa) I-84084, Italy\\
$^5$Kamerlingh Onnes Laboratory, Leiden University, P.O.B. 9504,
2300 RA Leiden, The Netherlands}

\date{\today}
%\maketitle

\begin{abstract}
Porous silicon, obtained by electrochemical etching, has been used
as a substrate for the growth of nanoperforated Nb thin films. The
films, deposited by UHV magnetron sputtering on the porous Si
substrates, inherited their structure made of holes of 5 or 10 nm
diameter and of 10 to 40 nm spacing, which provide an artificial
pinning structure. The superconducting properties were
investigated by transport measurements performed in the presence
of magnetic field for different film thickness and substrates with
different interpore spacing. Perpendicular upper critical fields
measurements present peculiar features such as a change in the
$H_{c2\bot}(T)$ curvature and  oscillations in the field
dependence of the superconducting resistive transition width at $H
\approx1$  Tesla. This field value is much higher than typical
matching fields in perforated superconductors, as a consequence of
the small interpore distance.
\end{abstract}


\maketitle

\section{Introduction}

In order to increase the critical parameters of both low- and
high-$T_{c}$ superconductors, in the last years great efforts were
directed towards the optimization of the vortex confinement in
these systems. The enhancement of the vortex pinning can be
obtained using periodic arrays of different types of artificial
defects in superconducting thin films, such as periodic variation
of thin film composition \cite{Raffy_SSC} or heavy ion irradiation
\cite{Harada}. Progress in the fabrication of nanostructures by
lithography has given access to a wide variety of well controlled
vortex regular pinning arrays at sub-micron length scales
\cite{Lykov,Baert,Morgan}. Experiments performed on this kind of
samples revealed the presence of peaks in the magnetic field
dependence of the critical current density $J_{c}(H)$
\cite{Fiory,Mosh_size,Castellanos} at matching fields
$H_{n}=nH_{1}$, where $n$ is an integer and $H_{1}$ is the value
of the field at which the vortex lattice spacing is equal to the
period of the regular array of submicrometer defects, \textit{a}:
$H_{1}=2\Phi_{0}/\sqrt{3}a^{2}$. Recently a radically different
approach to nanostructures fabrication based on self assembled
growth \cite{ChemRev}, such as chemically anisotropic etching of
single crystals, attracted much attention also in the
superconducting field. These processes are useful for generating,
on large areas, low-cost simple patterns of nanostructures in a
single step, assuring a high reproducibility as well. Moreover,
these techniques give the opportunity to produce regular periodic
structures with features that can be even an order of magnitude
smaller than the ones obtained by time consuming e-beam
lithography process, leading to templates having features
comparable with the characteristic superconducting lengths. In
particular, the reduced dimensions are of great importance in the
enhancement of the superconducting pinning properties, since
smaller interpore spacing implies higher matching fields down to
lower temperatures. At this purpose transport and magnetic
measurements were already performed on Nb thin films grown on
nanoporous Al$_{2}$O$_{3}$ templates with pores diameter in the
range 25-150 nm \cite{Prischepa,Welp,Vinckx1,Vinckx2,Vanacken}.
Anomalies in the $M(H)$ as well as in the temperature dependence
of the perpendicular upper critical field $H_{c2\bot}(T)$ were
observed.

The aim of this work is to present the superconducting properties
of Nb thin films deposited on porous silicon (PS) substrates
fabricated by electrochemical etching of monocrystalline Si in an
HF solution. Porous silicon is constituted by a network of pores
immersed in a nanocrystalline matrix \cite{Pavesi}. As a result,
PS is characterized by a very large chemically reactive internal
surface area, which makes this material promising for many
technological applications in different fields, as for instance
micro and optoelectronics, gas sensing, and biotechnology;
recently PS was also applied as a substrate for the nanotubes
growth \cite{Pavesi,Jakubowicz}. The reason to choose PS as a
template for superconducting thin films deposition is twofold. On
one hand Si based substrates are good candidates for the Nb
growth. Moreover, the characteristic pore features of these PS
substrates are the smallest used in superconducting field at this
purpose. The porous area of our substrates extends, in fact, on
about 1 cm$^{2}$ and it is constituted by pores of mean diameter,
${\O}$, tunable between 5 and 10 nm, with interpore spacing (the
distance between the centers of two nearest pores), $\Lambda$,
that can be varied from 10 to 40 nm. It is worth to underline that
the regularity of the pores arrangement in PS substrates is lower
than the one observed in Al$_{2}$O$_{3}$ templates obtained by
electrochemical oxidation \cite{Prischepa2,MTemplfi,Prischepa3}.
However it has been recently shown that the interaction of the
vortex lattice with templates of artificial pinning centers
presenting a short-range order can also give rise to peculiar
behavior of both critical magnetic fields and critical currents
\cite{Eisenmenger}. In the same way, thin Nb films deposited on PS
substrates can inherit their structure, resulting in porous Nb
thin films with in plane geometrical dimensions, $\Lambda$ and
${\O}$, comparable with the superconducting coherence length,
$\xi(T)$. This reduced periodicity could reflect in peculiar
properties of porous thin films, that could be interesting also in
view of possible application. For instance tuning the samples
geometry ($\Lambda$, ${\O}$) we expect to be able to control the
superconducting critical temperature, to reach higher matching
fields and eventually also higher critical currents, down to lower
temperatures. The superconducting properties of porous Nb thin
films have been probed by transport measurements in presence of
magnetic fields applied in the direction perpendicular to the
samples surface. As a consequence of the high density of the pore
network, the (H,T) phase diagrams present a deviation from the
classical behavior of $H_{c2\bot}(T)$, and the $\Delta T_{c}(H)$
behavior shows fingerprints of the periodic anomaly typical of
superconducting wire networks. Both these effects appear at fields
up to $H \approx1$  Tesla. This value is larger than typical
matching fields of periodic pinning arrays obtained both by
lithographic techniques and by using self-organized
Al$_{2}$O$_{3}$ templates.

\section{Fabrication and characterization}

\subsection{Substrates}

Porous layers were fabricated by electrochemical anodic etching of
p-type, 12 $\Omega$cm, and n-type, 0.01 $\Omega$cm,
monocrystalline silicon wafers in 48\% water solution of HF at the
current density of 25 mA/cm$^{2}$. n-type silicon wafers were
anodized with light exposition by 100 W tungsten lamp as well
without light illumination. The anodization time was chosen in the
range of 0.5 - 4 min in order to get porous layers with a
thickness ranging from 0.5 to 4 $\mu$m. The integral porosity was
estimated by gravimetry to be in the range from 30\%  up to 60\%
\cite{Lazarouk}. For p-type silicon wafers the nominal pore
diameter and the interpore spacing are respectively ${\O}=5$ nm
and $\Lambda=10$ nm. In the case of n-type silicon wafers the
nominal pore diameter is ${\O}=10$ nm, while the values of
$\Lambda$ are expected to be 20 nm for light exposed substrates
and 40 nm for samples obtained without illumination.

\begin{figure*}
\includegraphics[width=12cm]{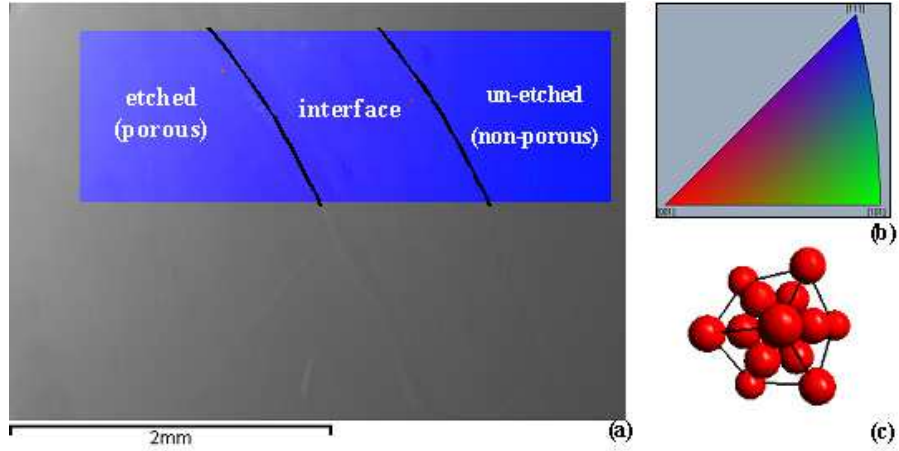}
\vspace{0.5in}
\caption{(color online). (a) Scanning electron microscope image
and EBSD orientation map of an n-type silicon wafer. The scanned
area crosses the interface between the etched (porous) and
un-etched (non-porous) part of the silicon wafer. The thick solid
lines in the blue area indicate the boundaries between different
regions. The color uniformity in the orientation map (blue area)
reveals that all of the crystalline components in this region have
the same orientation. From the color key (b) it emerges that the
crystal orientation of the analyzed area has the [111] direction
parallel to the wafer normal direction, as schematically shown in
(c). \label{EBSD}}
\end{figure*}

The Electron Back Scatter Diffraction (EBSD) technique has been
used to investigate the microstructure of both p-type and n-type
porous silicon wafers. EBSD patterns were obtained from LEO EVO 50
Scanning Electron Microscope with an Oxford Inca detector. In
order to optimize the contrast in the diffraction pattern and the
fraction of electrons scattered from the sample, the substrates
were tilted so that an angle of $70^{\circ}$ is formed between the
normal to the sample surface and the electron beam \cite{EBSD}.
EBSD patterns were acquired on porous and non-porous regions of
the two different substrates. Fig. \ref{EBSD} shows crystal
orientation map of a 3 mm by 0.5 mm area for a n-type substrate
with a spatial resolution of about 80 ${\mu}$m$^{2}$. The scanned
area was selected crossing the interface between the etched
(porous) and un-etched (non-porous) silicon wafer. The color
uniformity of the map reveals that in this area all the
crystalline components have (111) orientation. Similarly, the same
analysis performed on p-type substrate showed (100)
crystallographic orientation on both porous and non-porous region.
In both cases the etching process does not influence the
crystalline properties of the substrates.

\subsection{Nb films}

Nb thin films were grown on top of the porous Si substrates in a
UHV dc diode magnetron sputtering system with a base pressure in
the low $10^{-8}$ mbar regime and sputtering Argon pressure of
$1\times10^{-3}$ mbar. Reference films were grown in each
deposition run on standard non-porous Si(100) substrates. In order
to reduce the possible contamination of the porous templates, the
substrates were heated at $120^{\circ}$C for one hour in the UHV
chamber. The deposition was then realized at room temperature
after the cool off of the system. The films were deposited at
typical rates of 0.3 nm/s, controlled by a quartz crystal monitor
calibrated by low-angle reflectivity measurements. Since the
effect of the periodic template would be reduced when the film
thickness, $d_{Nb}$, exceedes the pore diameter, ${\O}$, the Nb
thickness was allowed to range between 8.5 and 15 nm. All the
samples deposited on porous Si, their names and thicknesses, and
the substrate characteristics as well, are summarized in Table
\ref{table}. For the sake of clarity the samples were named using
the initials Si followed by a number indicating the nominal
substrate interpore distance, $\Lambda$, and by another number for
the Nb thickness. For example, Si40-Nb12 is the Nb film 12 nm
thick, grown on the porous substrate with $\Lambda=40$ nm. The
pore diameter is in this case ${\O}=10$ nm.

\begin{figure*}
\includegraphics[width=12cm]{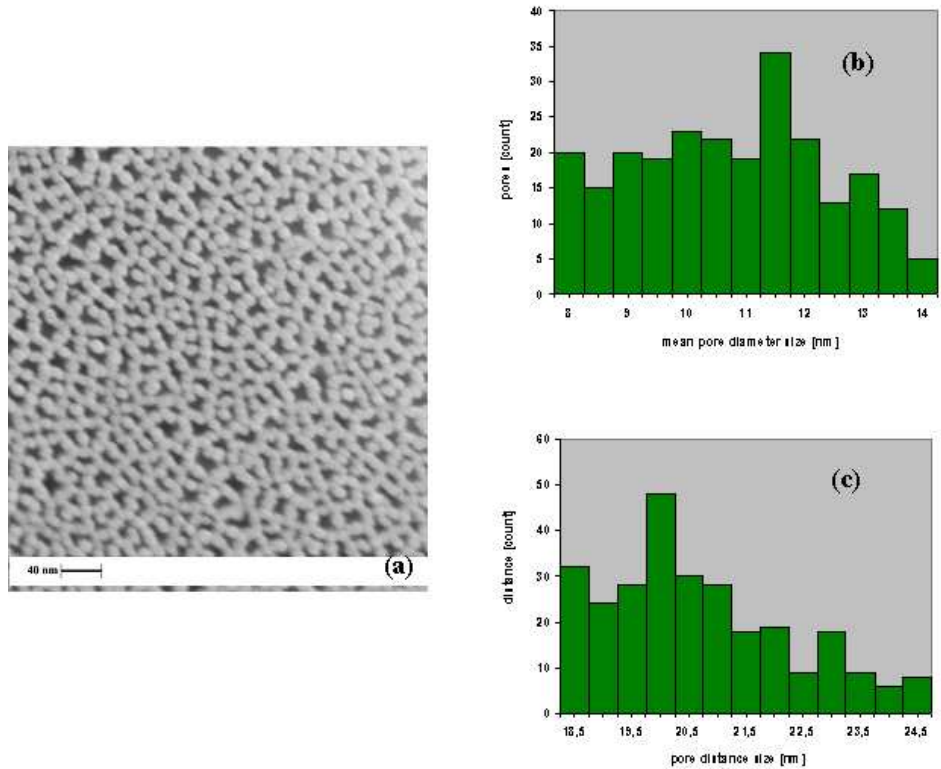}
\vspace{0.5in} \caption{(color online). (a) FESEM image
(magnification = 500000) of the top of Si20 substrate covered with
a conductive Pd/Au layer, 8 nm thick. Histogram of the pore
diameter (b) and pore spacing (c) distributions. \label{FESEM_Si}}
\end{figure*}

\begin{figure*}
\includegraphics[width=12cm]{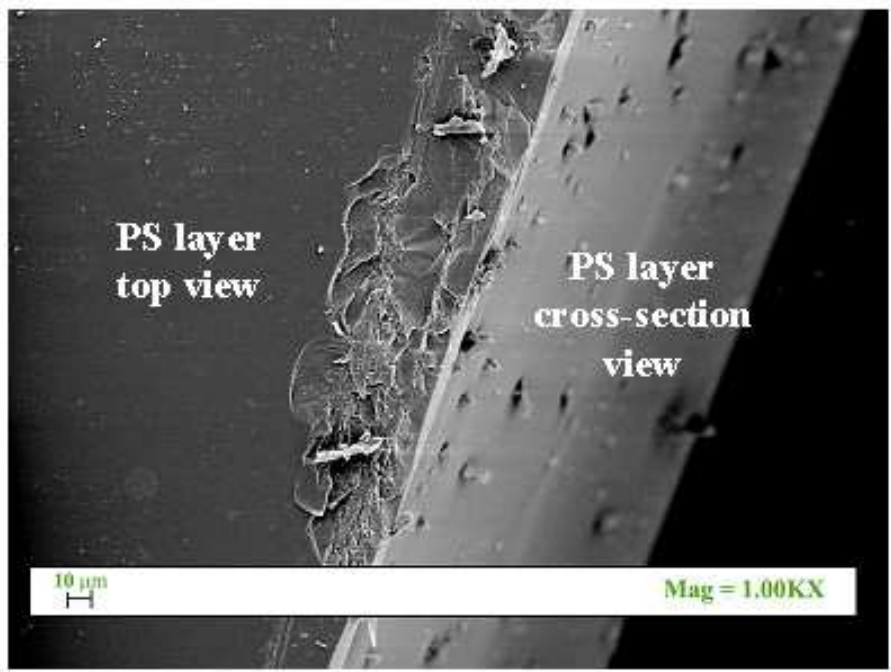}
 \vspace{0.5in} \caption{Low resolution FESEM image of the
edge of the Si20 substrate. \label{FESEM_low}}
\end{figure*}

\begin{figure*}
\includegraphics[width=12cm]{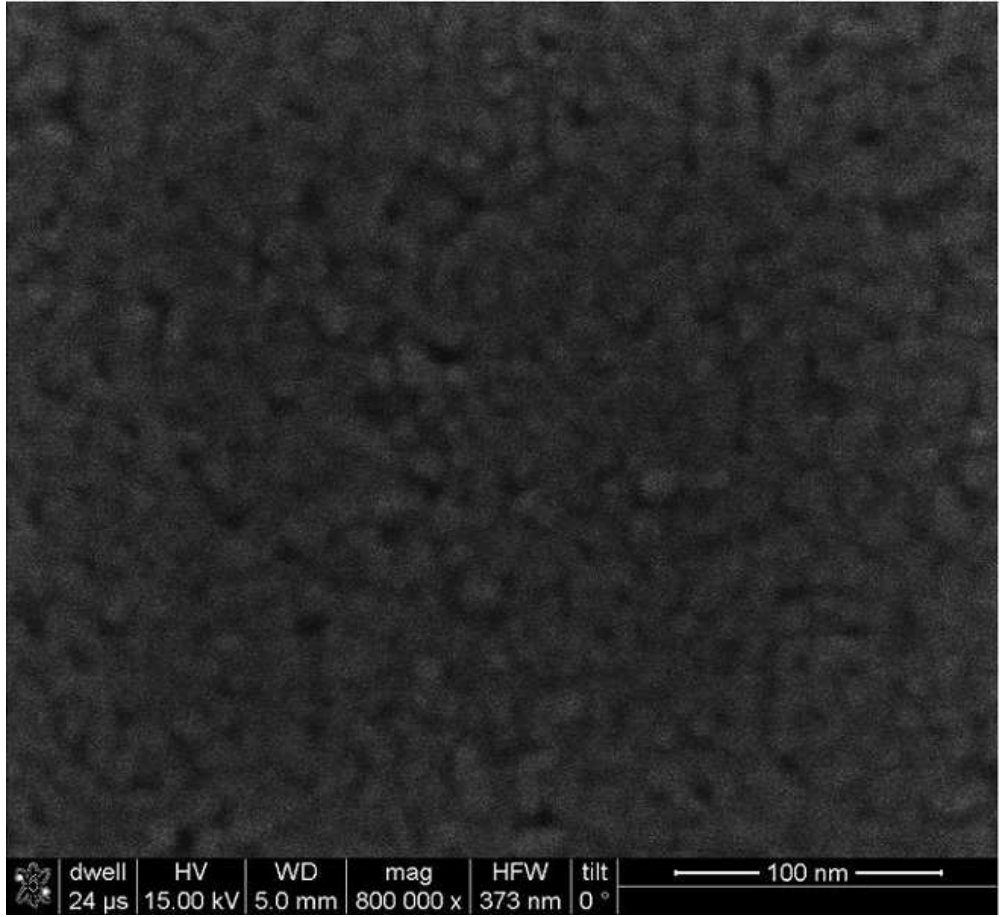}
\vspace{0.5in} \caption{FESEM image of a Nb film 9 nm thick
deposited on the PS Si20.} \label{FESEM_NL}
\end{figure*}

The morphology of the PS substrates as well as of the Nb thin
films deposited on them was analyzed by field emission scanning
electron microscopy (FESEM). To allow FESEM measurements on highly
resistive PS substrates, metallization with 8 nm Pd/Au thin film
was realized. A typical FESEM micrograph (magnification = 500000)
of the top of the Si20 substrate is shown in Fig.
\ref{FESEM_Si}(a). At this high magnification the pores can be
clearly distinguished as dark areas surrounded by Pd/Au molecular
agglomerates. Fig. 2a also reveals that the pore entrances are not
regular circles. Several micrographs obtained on PS templates were
analyzed in order to evaluate the pore diameter and the interpore
spacing distributions resulting from the electrochemical etching.
In order to take into account the pores shape we evaluated an
\lq\lq equivalent pore diameter\rq\rq, defined as the diameter of
a circle having the same surface area of the pore. The latter was
obtained by the help of ImageTool, a software for image analysis,
which enables to supply the surface area of a region drawing a
polygon around its outline. The evaluated pore size distribution
of the PS layer is reported in Figure 2b (mean diameter
${\O}=10.35$ nm and standard deviation $\sigma_{{\O}}=1.90$ nm).
Moreover, the mean interpore distance $\Lambda=20.42$ nm with a
standard deviation of $\sigma_{\Lambda}=1.80$ nm (Figure 2c) were
estimated.  In Fig. \ref{FESEM_low} a low resolution FESEM image
of the Si wafer edge is reported. Along this edge, at about 10
$\mu$m from the top surface, a change in the morphology can be
distinguished, indicating the presence of a porous layer of about
10 $\mu$m, in agreement with sample specifications. All these
values are in very good agreement with the expected nominal ones
and indicate the effectiveness of the porogenic treatment. In Fig.
\ref{FESEM_NL} a secondary electron FESEM image (magnification =
800000) of a dedicated Nb film, 9 nm thick, deposited on a Si20
substrate is shown. Again, the pores restrict the solid angle from
which low energy electrons can escape, thus rendering the pores
dark and the top surface light. The surface appearance, far from
being perfectly ordered, however clearly reveals the presence of
holes in the Nb layer.

\section{Superconducting properties}

The superconducting properties, critical temperatures $T_c$, and
perpendicular upper critical fields as a function of the
temperature, $H_{c2\bot}(T)$, were resistively measured in a
$^{4}$He cryostat using a standard dc four-probe technique on
unstructured samples. $T_c$ was taken at the 50\% of the
transition curves.

\begin{figure*}
\includegraphics[width=12cm]{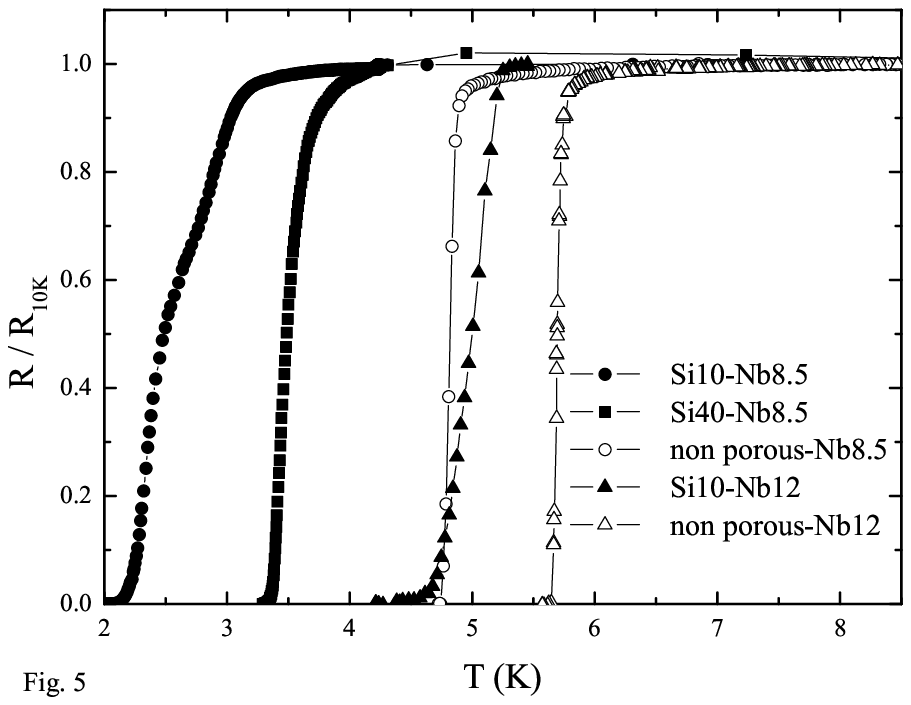}
\vspace{0.5in} \caption{Normalized $R(T)$ transition curves of Nb
thin films of different thickness grown on the same porous Si10
substrate, namely samples Si10-Nb8.5 and Si10-Nb12. For
comparison, the resistive transitions of non-porous reference Nb
films of the same thickness are shown. Also, the $R(T)$ curve is
presented for the sample Si40-Nb8.5. $R_{10K}$ is the resistance
value at T = 10 K. \label{RT}}

\end{figure*}

In Fig. \ref{RT} are reported the $R(T)$ transition curves of Nb
thin films of different thickness grown on the porous substrates
Si10, namely samples Si10-Nb8.5 and Si10-Nb12, and Si40, namely
sample Si40-Nb8.5. It can be observed that the critical
temperature, $T_{c}$, decreases as the thickness of the Nb film is
reduced, as expected \cite{Minhaj}. The same $T_{c}$($d_{Nb}$)
dependence is obtained for samples sets deposited on different
porous templates. In the same figure the resistive transitions of
non-porous reference Nb films are also shown. The critical
temperatures of these samples are higher than the $T_{c}$ of the
film with the same thickness deposited on the porous substrates.
This difference is less evident for the sample with the larger
interpore distance, since the ratio between the non-porous and the
porous area is higher in this case so that the critical
temperature results less depressed. This $T_{c}$ reduction could
be related, for example, to the highly reactive internal porous
surface area \cite{Pavesi}. From Fig. \ref{RT} it also emerges
that the $T_{c}$ difference between porous and non-porous samples
is reduced as the Nb thickness is increased due to the fact that,
for larger $d_{Nb}$, the pores are mostly covered and their effect
on the electronic transport properties is less important. The data
show also that the $R(T)$ shapes are strongly effected by the
pores and the interpore distances dimensions. In fact, while
sample Si40-Nb8.5 shows a rather sharp $R(T)$ curve, sample
Si10-Nb8.5 presents a broad double step transition. This
difference, at these reduced Nb thickness and substrate
parameters, can be ascribed to the formation of a superconducting
wire network made of several thin Nb stripes of different width,
and hence to the distribution of transition temperatures for the
stripes in the entire sample \cite{Pannetier2,Wilks,Zant,Rammal}.

The (H,T) phase diagrams for our samples were obtained performing
$R(T)$ measurements at fixed magnetic fields and $R(H)$
measurements at fixed temperatures. In general, the perpendicular
upper critical field of superconducting films of thickness
\textit{d} obeys to a linear temperature dependence,

\begin{equation}
H_{c2\bot}(T)=\;\frac{\Phi_{0}}{2\pi\xi_{0\parallel}^{2}}\;\Bigg(1-\frac{T}{T_{c}}\Bigg)
\label{eq:Hc2}
\end{equation}

\noindent where $\xi_{0\parallel}$ is the parallel Ginzburg-Landau
coherence length at T = 0. The temperature dependence of
$\xi_{\parallel}(T)$ is $\xi_{\parallel}(T)$ =
$\xi_{0\parallel}$/$\sqrt{1-T/T_{c}}$. Since the film dimensions
in the \textit{xy} plane are larger than $\xi_{\parallel}(T)$, the
expression (\ref{eq:Hc2}) is verified in the whole temperature
range.

\begin{figure*}
\includegraphics[width=12cm]{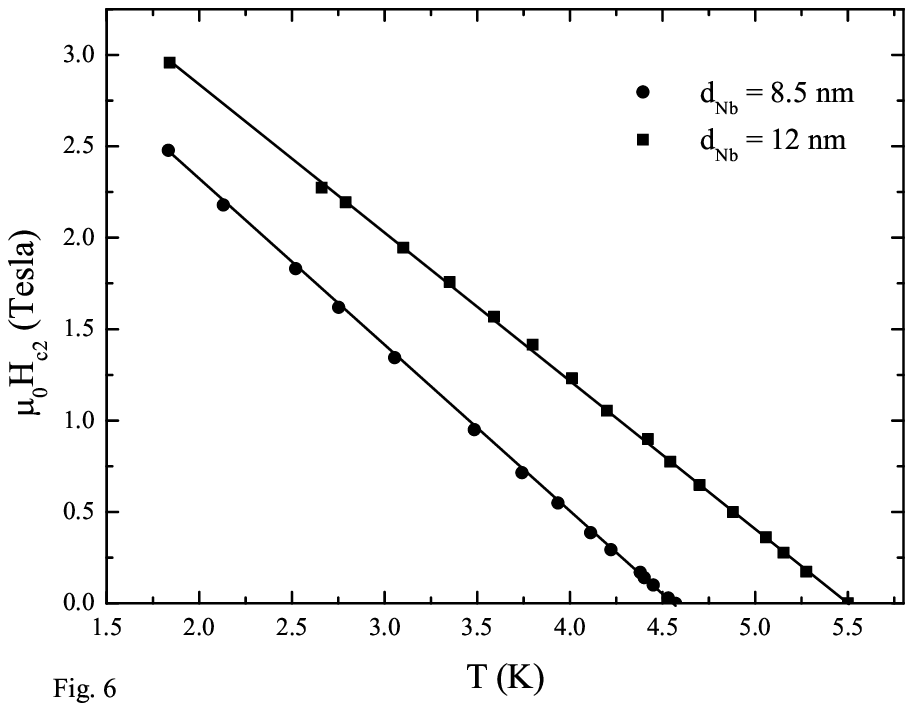}
\vspace{0.5in} \caption{Perpendicular upper critical field
$H_{c2\bot}$ vs. temperature of non-porous reference Nb films with
$d_{Nb}=8.5$ nm (circles) and $d_{Nb}=12$ nm (squares). The solid
lines represent the linear fit to the data. \label{HT-nPS}}

\end{figure*}

Before presenting the results obtained for the Nb films deposited
on PS, it is worth to show the $H_{c2\bot}(T)$ curves for two Nb
reference films, as reported in Fig. \ref{HT-nPS}. As expected the
$H_{c2\bot}(T)$ behavior is linear for both the samples. A fit of
the data with the expression (\ref{eq:Hc2}), yields a value of the
Ginzburg-Landau coherence length at T = 0, $\xi_{0\parallel}=9.5$
nm for $d_{Nb}=8.5$ nm and $\xi_{0\parallel}=8.7$ nm for
$d_{Nb}=12$ nm. Thus the pore diameter dimensions in our PS
substrates are comparable with the vortex core dimensions at T=0,
${\O}\approx\xi_{0\parallel}$. This means that each pore can trap
only one fluxon, the saturation number being
$n_{s}={\O}/2\xi(T)\leq1$ \cite{Mkrtchyan}. Subsequently
multiquanta vortex lattice \cite{Mosh_size} can not be observed in
our systems, which are then suitable for the study of
commensurability effect at low temperatures and high magnetic
fields.

\begin{figure*}
\includegraphics[width=8cm]{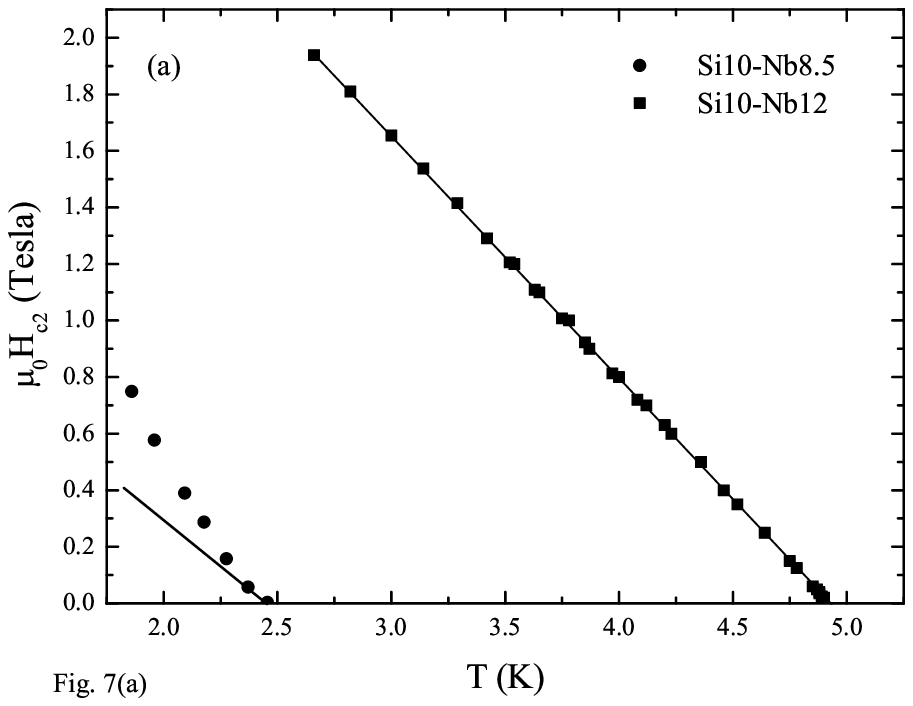}
\includegraphics[width=8cm]{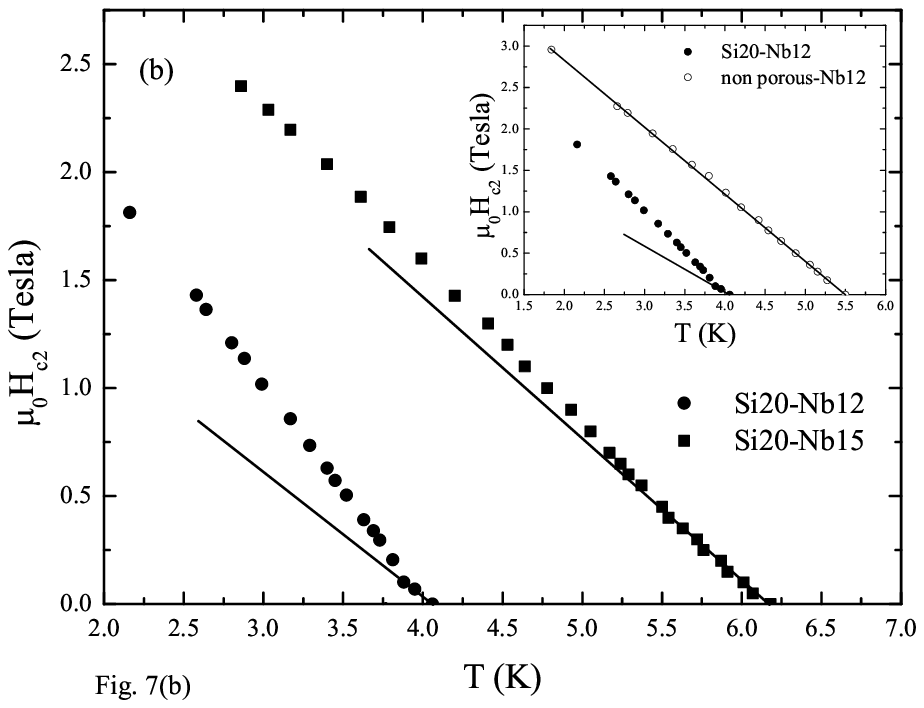}
\vspace{0.5in}\caption{(a) Superconducting phase diagram (H,T) of
Nb thin films of different thickness grown on the porous Si10
substrate, namely samples Si10-Nb8.5 and Si10-Nb12. The linear
fits of the data of the samples Si10-Nb8.5 and Si10-Nb12 are shown
in figure: near $T_{c}$ for the sample Si10-Nb8.5 and on whole
temperature range for the second one. (b) $H_{c2\bot}(T)$
dependence of two samples of different thickness grown on the
porous Si20 template, namely samples Si20-Nb12 and Si20-Nb15. The
linear fits of the data of the samples Si20-Nb12 and Si20-Nb15
near $T_{c}$ are shown in figure. The inset shows the comparison
among two samples of the same thickness, grown on the porous
template Si20 and on the non-porous template.\label{HT-PS1}}
\end{figure*}

The $H_{c2\bot}(T)$ curves obtained for the Nb films deposited on
porous Si templates present some peculiarities. Fig.
\ref{HT-PS1}(a) shows the phase diagram, $H_{c2\bot}(T)$, of two
samples of different thickness grown on the porous Si substrate
with $\Lambda=10$ nm and ${\O}=5$ nm, namely Si10-Nb8.5 and
Si10-Nb12. Similarly, Fig. \ref{HT-PS1}(b) shows $H_{c2\bot}(T)$
dependence of two samples of different thickness grown on the
porous Si substrate with $\Lambda=20$ nm and ${\O}=10$ nm, namely
Si20-Nb12 and Si20-Nb15. In both cases it is clearly visible that
the introduction of the porous array modifies the upper critical
field behavior of the Nb thin film. $H_{c2\bot}(T)$ is
characterized, in fact, by a strong non linear temperature
dependence near $T_{c}$, which becomes less evident when the Nb
thickness exceeds ${\O}$. In the sample Si10-Nb12, where the
condition $d_{Nb}=12$ nm $\geq {\O}=10$ nm is realized, the upper
critical magnetic field is linear in all the temperature range, as
shown by the linear fit in Fig. \ref{HT-PS1}(a). The comparison
between two Nb thin films of the same thickness deposited on Si20
and a non-porous substrate is reported in the inset of Fig.
\ref{HT-PS1}(b). Apart from a $T_{c}$ reduction the porous Nb film
is again characterized by a non linear $H_{c2\bot}(T)$ dependence.
However, for these samples the positive curvature of
$H_{c2\bot}(T)$ close to $T_{c}$ is the only effect we could
notice since the extremely small interpore spacing of these two
analyzed substrates (Si10, Si20) would produce commensurability
effects only at very large magnetic fields, namely 24 and 6 Tesla,
respectively. These fields values are not accessible to our
experimental set up. On the other hand the matching field should
be considerably smaller if Nb films are deposited on a PS
substrate with a larger interpore distance. For example, for the
porous Si template Si40, $H_{1}\approx1.4$ Tesla, a reachable
value for our experimental set up. For this reason, to study
commensurability effects, two films of different thickness were
deposited on this substrate, namely samples Si40-Nb8.5 and
Si40-Nb12. Moreover, in order to investigate a film with an
intermediate Nb thickness, the sample Si40-Nb12 was deliberately
oxidized obtaining the formation of about 2 nm thick oxide layer.
This sample, with a resulting suppressed superconducting critical
temperature, was named Si40-Nb10. The $R(T)$ transitions measured
for different values of the magnetic fields from 0 to 2 Tesla are
presented in the inset of Fig. \ref{HT-PS2}(a) for the sample
Si40-Nb10. The resulting (H,T) phase diagram is shown in Fig.
\ref{HT-PS2}(a). Similarly, in Fig. \ref{HT-PS2}(b) the
$H_{c2\bot}(T)$ dependence is shown for the sample Si40-Nb12. The
upper critical fields are again characterized by a non linear
temperature dependence near $T_{c}$.

Moreover, it is also interesting to note that for these samples
the $H_{c2\bot}(T)$ dependencies present first a positive
curvature followed by a change in the concavity, which occurs in
both cases at $H \approx1$  Tesla. This effect is more evident if
we plot the $H_{c2\bot}$ second derivative versus temperature as
shown in Fig. \ref{HT-PS2}(a),(b), right scale. In both cases
$dH_{c2\bot}^{2}$/$dT^{2}$ shows, in fact, an abrupt change at $H
\approx1$ Tesla, where the anomaly in the $H_{c2\bot}(T)$
dependence is present. Similar features in (H,T) phase diagram
were reported for Nb thin films deposited on Al$_{2}$O$_{3}$
substrates \cite{Prischepa,Welp,Vinckx1}, as well as on triangular
array produced by micellar technique \cite{Eisenmenger}. Another
peculiarity that can be ascribed to the reduced dimensionality of
these samples is observed plotting the $R(T)$ transition width,
$\Delta T_{c}$ = $T_{c}$($R$ = $90\%$ $R_{10K}$)$-T_{c}$($R$ =
$10\%$ $R_{10K}$), as a function of the field. The $\Delta T_{c}$
dependence, reported in inset of Fig. \ref{HT-PS2}(b) for the
sample Si40-Nb12 shows, in fact, a minimum again at $H \approx1$
Tesla. This feature is in general expected at the matching field,
since dissipative effects are reduced as commensurability occurs,
resulting in shaper resistive transitions. Both anomalies of the
$H_{c2\bot}(T)$ and the $\Delta T_{c}(H)$ behaviors at $H
\approx1$ Tesla, make us confident to identify this field as the
first matching field, $H_{1}$. Then, it follows that the period of
the porous template is \textit{a} = 48 nm, a value in very good
agreement with the nominal interpore spacing of this analyzed
sample, $\Lambda=40$ nm. The non-monotonic $\Delta T_{c}(H)$
behavior is also reminiscent of the formation of a superconducting
wire network. In this case the observed oscillations could be a
direct consequence of fluxoid quantization in a multiply connected
superconductor, an extension of Little-Parks oscillations of an
individual superconducting loops \cite{Parks}. The expected
amplitude of the $\Delta T_{c}$ oscillations can be estimated by
the expression $T_{c} [\xi_{0}/a]^{2}$ \cite{Pannetier}, which in
our case, for \textit{a} = 48 nm, gives $\Delta T_{c}=180$ mK, a
value close to the one experimentally observed, namely $\Delta
T_{c} \approx120$ mK.

A fit of the data close to $T_{c}$ with the expression
(\ref{eq:Hc2}), yields a value of the Ginzburg-Landau coherence
length at T = 0, $\xi_{0\parallel}=9.4$ nm and
$\xi_{0\parallel}=8.7$ nm, respectively. The values of
$\xi_{0\parallel}$ are significantly smaller than the BCS
coherence length of Nb, $\xi_{0}=39$ nm \cite{Buckel}, indicating
that our films are in dirty limit regime with an electron mean
free path of $\textit{l}$ = 1.38 $\xi_{0\parallel}^{2}$ /
$\xi_{0}$ $\approx3$ nm \cite{Schmidt}. The Ginzburg-Landau
parameter, $\kappa$ = $\lambda$(0)/$\xi_{0\parallel}$, can be
estimated using the expression $\kappa$ =
0.72$\lambda_{L}$/$\textit{l}$ $\approx10$, where $\lambda_{L}=39$
nm is the London penetration depth of Nb \cite{Buckel}. Ratios of
$\xi_{0\parallel}$/$\textit{a} \approx0.2$ and
$\lambda$(0)/$\textit{a} \approx2.0$, measured for \textit{a} = 48
nm, are larger than in previous works \cite{Welp,Vinckx1} on
perforated Nb samples, indicating that we are in presence of
individual vortex pinning \cite{Brandt}. Therefore the porous
template can overcome the lattice rigidity, being suitable for the
vortex lattice pinning in Nb thin films.

\section{Conclusions}

Critical temperatures and perpendicular upper critical fields
measurements were performed on Nb thin films sputtered on
different PS substrates obtained by electrochemical etching of Si
in a HF solution. The control of the etching parameter allows to
obtain templates with pores diameter and interpore distance
tunable in the range ${\O}=5-10$ nm and $\Lambda=10-40$ nm,
respectively. These templates were used as a system of pinning
arrays in order to obtain the formation of commensurate vortex
structures at high matching fields down to low temperatures.
Anomalies in the $H_{c2\bot}(T)$ behavior, as well as in the field
dependence of the $R(T)$ width, were in fact observed at $H
\approx1$ Tesla, which was estimated as the first matching field,
$H_{1}$. These preliminary results are promising, specially
considering that PS prepared by traditional electrochemical
etching exhibits a sponge structure, so that the holes are not a
regularly distributed ordered planar pinning structure. Future
work will focus on critical currents density studies, carried out
both by transport measurements on patterned samples and by
magnetization measurements. Moreover, due to the extreme reduced
features size of PS templates, these systems look appealing also
for the studying of two dimensional superconducting wire networks.

\vspace{0.5in}

S.L. Prischepa gratefully acknowledges the CNR-Italy for financial
support within the CNR Short Term Mobility Program.

\newpage

\begin{table}
\vspace{0.5in} \caption{Characteristics of porous Si templates and
properties of Nb thin films deposited on such substrates.
$\Lambda$ indicates the interpore spacing, ${\O}$ the pore
diameter, $d_{Nb}$ the Nb thickness and $T_{c}$ the critical
temperature of the samples. The interpore spacing and the pore
diameter are respectively $\Lambda=10$ nm and ${\O}=5$ nm for
p-type silicon wafers, and $\Lambda=20$ nm up to 40 nm and
${\O}=10$ nm for n-type silicon wafers.}

\vspace{4mm}

\begin{tabular}{ccccc}

Sample  & ${\O}$ (nm) & $\Lambda$ (nm) & $d_{Nb}$ (nm) & $T_{c}$ (K) \\
\hline
Si10-Nb8.5 & 5$\pm$1  & 10$\pm$2  & 8.5  & 2.48   \\

Si10-Nb12  & 5$\pm$1  & 10$\pm$2  & 12   & 4.68   \\

Si20-Nb12  & 10$\pm$2   & 20$\pm$4  & 12   & 4.06   \\

Si20-Nb15  & 10$\pm$2   & 20$\pm$4  & 15   & 6.18   \\

Si40-Nb8.5 & 10$\pm$2   & 40$\pm$6  & \textbf{8.5}   & 3.49   \\

Si40-Nb10 & 10$\pm$2   & 40$\pm$6  & 10   & 5.17   \\

Si40-Nb12  & 10$\pm$2   & 40$\pm$6  & 12   & 5.94   \\
\end{tabular}
\label{table}
\end{table}

\newpage

\newpage
\newpage

\newpage

\newpage
\newpage

\newpage

\newpage

\begin{figure*}
\includegraphics[width=12cm]{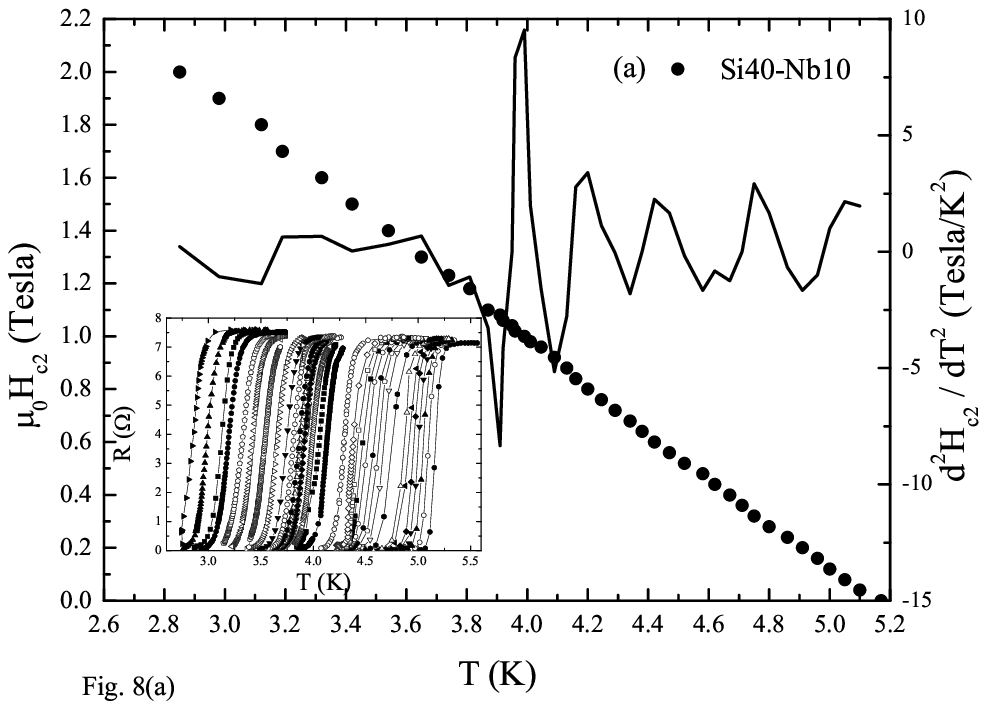}
\includegraphics[width=12cm]{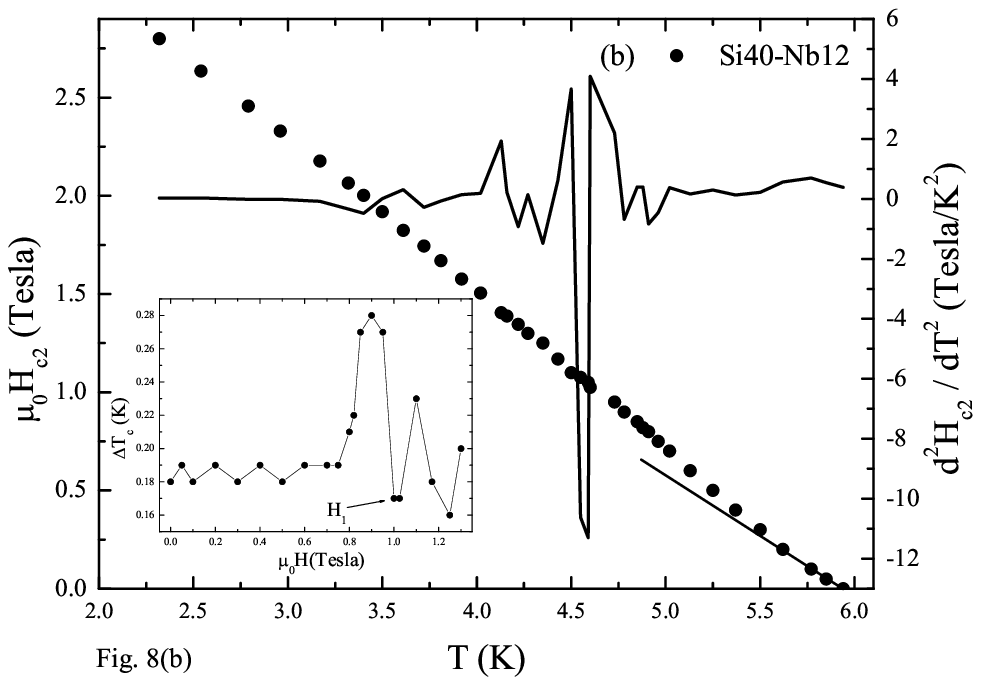}
\end{figure*}

\begin{figure*} \vspace{0.5in} \caption{(a) Left scale:
Superconducting phase diagram (H,T) of the sample Si40-Nb10. The
solid line represents the linear fit to the data near $T_{c}$.
Right scale: $dH_{c2\bot}^{2}$/$dT^{2}$ versus temperature is
presented. The inset shows resistive transitions of the same
sample, measured for different values of the magnetic fields, from
0 to 2 Tesla. (b) Left scale: (H,T) phase diagram of the sample
Si40-Nb12. The solid line represents the linear fit to the data
near $T_{c}$. Right scale: $dH_{c2\bot}^{2}$/$dT^{2}$ versus
temperature is shown. The inset presents the $R(T)$ transition
width, $\Delta T_{c}$ = $T_{c}$($R$ = $90\%$
$R_{10K}$)$-T_{c}$($R$ = $10\%$ $R_{10K}$), as a function of the
field. The first matching field is indicated.\label{HT-PS2}}
\end{figure*}

\end{document}